\documentclass[conference]{IEEEtran}
\usepackage{cite}
\usepackage{amsmath,amssymb,amsfonts}
\usepackage{graphicx}
\usepackage{textcomp}
\usepackage{xcolor}

\usepackage{tikz}
\usepackage{float}
\usepackage{adjustbox}
\usepackage{wrapfig}
\usepackage{placeins}
\usepackage{siunitx}
\usepackage{algorithm}
\usepackage[noend]{algpseudocode}
\usepackage{multirow}
\usepackage{refcount}
\usepackage{tablefootnote}
\usepackage[hidelinks]{hyperref}
\def\BibTeX{{\rm B\kern-.05em{\sc i\kern-.025em b}\kern-.08em
    T\kern-.1667em\lower.7ex\hbox{E}\kern-.125emX}}
\begin{document}

\title{Accelerating Clique Counting in Sparse Real-World Graphs via Communication-Reducing Optimizations}

\author{\IEEEauthorblockN{Amogh Lonkar}
% \IEEEauthorblockA{\textit{Baskin School of Engineering} \\
\IEEEauthorblockA{\textit{Computer Science \& Engineering} \\
\textit{University of California, Santa Cruz}\\
Santa Cruz, CA, USA \\
alonkar@ucsc.edu}
\and
\IEEEauthorblockN{Scott Beamer}
% \IEEEauthorblockA{\textit{Baskin School of Engineering} \\
\IEEEauthorblockA{\textit{Computer Science \& Engineering} \\
\textit{University of California, Santa Cruz}\\
Santa Cruz, CA, USA \\
sbeamer@ucsc.edu}
}
\maketitle

\begin{abstract}
Counting instances of specific subgraphs in a larger graph is an important problem in graph mining. Finding cliques of size k (k-cliques) is one example of this NP-hard problem. Different algorithms for clique counting avoid counting the same clique multiple times by pivoting or ordering the graph. Ordering-based algorithms include an ordering step to direct the edges in the input graph, and a counting step, which is dominated by building node or edge-induced subgraphs. Of the ordering-based algorithms, kClist is the state-of-the art algorithm designed to work on sparse real-world graphs. Despite its leading overall performance, kClist's vertex-parallel implementation does not scale well in practice on graphs with a few million vertices.

We present CITRON (Clique counting with Traffic Reducing Optimizations) to improve the parallel scalability and thus overall performance of clique counting. We accelerate the ordering phase by abandoning kClist's sequential core ordering and using a parallelized degree ordering. We accelerate the counting phase with our reorganized subgraph data structures that reduce memory traffic to improve scaling bottlenecks. Our sorted, compact neighbor lists improve locality and communication efficiency which results in near-linear parallel scaling. CITRON significantly outperforms kClist while counting moderately sized cliques, and thus increases the size of graph practical for clique counting.
% allowing for faster clique counting in practice.

We have recently become aware of ArbCount \cite{shi2021parallel}, which often outperforms us. However, we believe that the analysis included in this paper will be helpful for anyone who wishes to understand the performance characteristics of k-clique counting.
\end{abstract}
 % as seen in Table \ref{table:3}

% kClist uses a sequential core ordering to convert the undirected input graph into a Directed Acyclic Graph (DAG), which takes a significant fraction of the total execution time. Furthermore, its existing  of kClist does not scale well in practice during the counting phase on graphs with a few million vertices.


\section{Introduction}
Clique finding is a well-researched topic in the graph community and it has many interesting practical applications for community detection \cite{6249683, palla2005uncovering, 7117352, fang2019efficient} and social network analysis \cite{4811845, rossi2015parallel}. In recent years, the data mining community has incorporated clique finding into deep learning classifiers to enhance recommender systems in social networks \cite{dietrec, movierec}. Clique finding is used prominently in bioinformatics, and researchers have used graph models to find variants in gene sequences \cite{variant}, efficiently group related genes in a database \cite{database} and perform protein structure analysis \cite{ppi}. The rapid growth in the number of social media users and the large size of genome data has amplified the need for high performance systems capable of analyzing these huge networks to count or enumerate their k-cliques.

Clique finding represents only a single problem in Graph Pattern Mining (GPM). Motifs are other general graph patterns that also interest researchers in this space. Various GPM frameworks have been developed to count repeating instances of given patterns within the search graph. These include Pangolin \cite{chen2020pangolin}, Peregrine \cite{jamshidi2020peregrine}, Arabesque \cite{teixeira2015arabesque}, Fractal \cite{dias2019fractal}, and others. The aforementioned frameworks use more generalized algorithms and provide their own APIs for counting instances of arbitrary, user-given patterns. In contrast, there are also algorithms specialized for counting of k-cliques, such as kClist \cite{danisch2018listing}, Pivoter \cite{jain2020power}, and Arb-Count \cite{shi2021parallel}. By virtue of being custom designed to solve a specific problem, these algorithms generally perform better than the general-purpose GPM frameworks.

KClist~\cite{danisch2018listing} is the current state-of-the-art clique counting algorithm, as it is algorithmically efficient and significantly outperforms prior work. However, when applied to large graphs, it demonstrates poor parallel scalability, especially for modest clique sizes. We carefully analyze kClist and identify scaling bottlenecks in both of its main phases: ordering \& counting. The ordering phase directionalizes the graph in order to avoid redundantly counting the same cliques multiple times in the subsequent counting phase. KClist uses a core ordering~\cite{matula1983smallest} to reduce the maximum degree in order to obtain the greatest reduction in the amount of algorithmic work in the counting phase. Since the ordering phase consumes a substantial portion of the execution time, and computing a core ordering is inherently sequential, it limits scalability due to Amdahl's Law. The parallel speedup of kClist's counting phase quickly saturates, and we observe speedups as poor as 5$\times$ on 24 cores for certain graphs. We find the counting phase is generally memory-bound due to poor locality and the sheer volume of memory traffic.

We present CITRON (Clique Counting with Traffic Reducing Optimizations), a high performance clique counting algorithm. Due to its better parallel scalability, it significantly outperforms kClist, especially for modest clique sizes. We find the use of a core ordering unnecessary, and a parallelizable ordering such as a degree ordering not only greatly accelerates the ordering phase, but it also results in negligible slowdowns in the counting phase in practice. We accelerate the counting phase with our novel subgraph construction approach which materially reduces the amount of memory accesses and improves their spatial locality.

We evaluate CITRON on a suite of real-world graphs and compare it to kClist and other prior works. For counting triangles, we achieve speedups of $18.16 - 208.27\times$ (Geometric Mean: $102.436\times$) in the ordering phase and speedups of $1.1-2.6\times$ (Geometric Mean: $1.798\times$) in the counting phase, resulting in total speedups of $14.37-39.37\times$ (Geometric Mean: $18.517\times$) over kClist.

We have recently become aware of ArbCount \cite{shi2021parallel}, which outperforms us as seen in Table \ref{table:3}. However, we believe that the analysis included in this paper will be helpful for anyone who wishes to understand the performance characteristics of k-clique counting.

\section{Background}

For a given undirected input graph \emph{G}, we want to count the number of cliques present of a given size k. A \emph{clique} is a completely connected subgraph, i.e. each vertex is connected to every other vertex in the subgraph. The input graph consists of a vertex set, \emph{V(G)}, and an edge set, \emph{E(G)}. For a k-clique \emph{C} present in the input graph, \emph{V(C) $\in$ V(G)}, \emph{E(C) $\in$ E(G)} and \emph{$\vert$V(C)$\vert$ = k}. Each vertex \emph{u} in \emph{G} has a neighborhood, which is the set of the vertices with which \emph{u} shares an edge. The neighborhood of \emph{u} in \emph{G} is indicated by \emph{N(u)} and its size is the degree $d(u) = |N(u)|$. We treat \emph{G} as an undirected graph, so $\forall$ \emph{v $\in$ N(u)}, then \emph{u $\in$ N(v)}.

To reduce the amount of work done when counting cliques, ordering-based algorithms transform \emph{G} into a directed acyclic graph (DAG), which we denote by $\vec{G}$. This step adds work to direct edges to remove cycles in order to avoid counting the same clique multiple times. Consequently, any vertex \emph{u} in the new DAG can have two types of neighbors: in-neighbors and out-neighbors. In-neighbors are those vertices in \emph{N(u)} for which \emph{u} is the destination vertex of the shared edge. Conversely, \emph{u's} out-neighbors are those vertices in \emph{N(u)} where \emph{u} is the source vertex of the edge. While counting cliques, we only consider the out-neighbors of a vertex \emph{u} in $\vec{G}$, denoted by $\vec{N(u)}$. The number of out-neighbors of a vertex \emph{u} is called its out-degree and is written as $\vec{d(u)}$.        

\begin{figure}
	\centering
	\begin{tikzpicture}[node distance={15mm}, thick, main/.style = {draw, circle}, scale=0.7, every node/.append style={transform shape}]
		\node[main] (1) {0};
		\node[main] (2) [right of=1] {1};
		\node[main] (3) [right of=2] {2};
		\node[main] (4) [below of=1] {3};
		\node[main] (5) [right of=4] {4};
		\node[main] (6) [right of=5] {5};
		\node [right=0.55cm] at (6) {$\longrightarrow$};
		\node[main] (7) [below of=5] {6}; 
		\draw (1) -- (2);
		\draw (1) -- (4);
		\draw (1) -- (5);
		\draw (2) -- (4);
		\draw (2) -- (5);
		\draw (2) -- (3);
		\draw (3) -- (6);
		\draw (4) -- (5);
		\draw (5) -- (7);
		\draw (5) -- (6);
		\draw (6) -- (7);
	\end{tikzpicture}
	\begin{tikzpicture}[node distance={15mm}, thick, main/.style = {draw, circle}, scale=0.7, every node/.append style={transform shape}] 
		\node[main] (1) {0};
		\node[main, red] (2) [right of=1] {1};
		\node[main] (3) [right of=2] {2};
		\node[main, red] (4) [below of=1] {3};
		\node[main, red] (5) [right of=4] {4};
		\node[main] (6) [right of=5] {5};
		\node[main] (7) [below of=5] {6}; 
		\draw[->] (1) -- (2);
		\draw[->] (1) -- (4);
		\draw[->] (1) -- (5);
		\draw[thick, red, ->] (4) -- (2);
		\draw[thick, red, ->] (2) -- (5);
		\draw[->] (3) -- (2);
		\draw[->] (3) -- (6);
		\draw[thick, red, ->] (4) -- (5);
		\draw[->] (6) -- (5);
		\draw[->] (7) -- (5);
		\draw[->] (7) -- (6);
	\end{tikzpicture}
	\caption{Converting an undirected input graph (left) to directed acyclic graph (right) by a degree-based ordering. Furthermore, the highlighted (red) portion on the right indicates the subgraph induced by vertex 0.}
	\label{exampleGraph}
\end{figure}
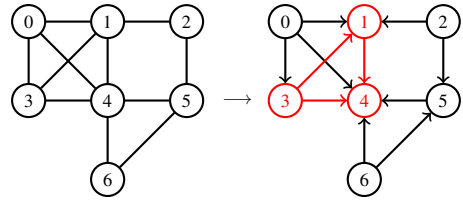

% There are various ways to produce an ordering for directionalizing the graph.
Given a total ordering $\omega$, \emph{directionalizing} transforms the graph \emph{G} to $\vec{G}$ by removing the edge \emph{u} $\rightarrow$ \emph{v} from \emph{E(G)} if \emph{$\omega$(u)} $\geq$ \emph{$\omega$(v)} and keeping only the edge \emph{v} $\rightarrow$ \emph{u} in \emph{E($\vec{G}$)}. Edges are thus directed from a lower $\omega$ to a higher $\omega$ vertex. kClist uses a core ordering, which guarantees the lowest maximum out-degree of a vertex in $\vec{G}$. While this approach reduces the amount of work done while counting cliques, it requires a fair bit of effort to compute, and cannot be parallelized~\cite{matula1983smallest}. Alternatively, a degree ordering uses the degree to compare vertices and uses the identifier as a tie breaker (example in Figure~\ref{exampleGraph}). We describe the degree ordering as:\\
$\omega: \omega(u) > \omega(v)$ if $(\vec{d(u)} > \vec{d(v)}) \lor (\vec{d(u)} = \vec{d(v)}) \land (u > v)$
% $\omega$: \emph{$\omega$(u)} $>$ \emph{$\omega$(v)} if $(\vec{d(u)} > \vec{d(v)}) \lor (\vec{d(u)} = \vec{d(v)} \land (u > v))$
% \\ Figure \ref{exampleGraph} demonstrates how an undirected graph is transformed into a DAG via the degree-based ordering.

A major step in counting cliques is building a vertex-induced subgraph. The induced subgraph contains the vertices in the neighborhood of the target vertex and any edges between them. The induced subgraph does not include the target vertex itself. The highlighted portion in Figure \ref{exampleGraph} shows the vertex 0 induced subgraph in the example graph. We denote the subgraph induced by vertex $u$ on $\vec{G}$ to be $\vec{g_u}$ with $\vec{V(g_u)} = N(\vec{u})$ and $\vec{E(g_u)} = \{ (v_1, v_2) \vert (v_1, v_2) \in E(\vec{G}) \land v_1 \in N(\vec{u}) \land v_2 \in N(\vec{u})\}$.
\section{Related Work}
One of the fastest sequential algorithms for clique counting is provided by Chiba and Nishizeki \cite{chiba1985arboricity}. First, it ranks the vertices in descending order of degree. Next, it searches the induced subgraph of each vertex recursively \emph{k-2} times where \emph{k} is the size of the desired clique. Once the recursion is complete, the vertex is removed from the graph to avoid duplicate work. The execution time of this algorithm is $O(kma(G)^{k-2})$, where \emph{m} is the number of edges in the input graph and \emph{a(G)} is the arboricity, i.e. the minimum number of forests into which the edges can be partitioned. While the algorithm is simple and efficient, the process of removing vertices from the graph makes it sequential and suboptimal for dealing with the massive real-world graphs of today. 

In recent years, more work has been done to parallelize clique counting to boost performance. Danisch et al. present kClist \cite{danisch2018listing}, the current state-of-the art parallel algorithm for counting and listing cliques. KClist is able to parallelize the counting step by first converting the input graph into a directed acyclic graph (DAG). Directionalizing the graph ensures kClist counts each clique only once without resorting to sorting the vertices and then removing them one by one sequentially while counting like Chiba and Nishizeki's algorithm. KClist uses a core ordering to rank vertices for directionalizing the graph to be used in the counting phase. The algorithm for computing the core ordering from Matula and Beck\cite{matula1983smallest} guarantees the smallest maximum out-degree in the graph. This is significant, because the maximum degree determines the execution time of the counting phase and a smaller value means less algorithmic work and a more efficient algorithm.

% In recent years, more work has been done to parallelize clique counting to boost performance. Danisch et al. developed kClist \cite{danisch2018listing}, the current state-of-the art parallel algorithm for counting and listing cliques. kClist converts the input graph into a directed, acyclic graph (DAG) instead of sorting the vertices and then removing them one by one. Operating on a DAG ensures that each clique is counted only once, and it removes the sequential step from Chiba and Nishizeki's algorithm. Different vertices can be processed in parallel to improve the total execution time. kClist uses a core ordering to rank vertices and generate the DAG (i.e. directionalize the graph) used in the counting phase. The algorithm for computing the core ordering from Matula and Beck\cite{matula1983smallest} guarantees the smallest maximum out-degree in the graph. This is significant, because the maximum degree determines the execution time of the counting phase and a smaller value means less algorithmic work and a more efficient algorithm.
% \\kClist computes its ordering in the following manner:

\begin{algorithm}[H]
	\caption{Core Ordering}\label{orderingAlg}
	\begin{algorithmic}[1]
		\Function{OrdCore}{G}
		\State $r \gets 0, n \gets \vert V(G) \vert$
		\State $R \gets [0, 0, ..., 0]$ \Comment{Array to hold rank of each node}
		\While{$n \neq 0$}
		\State Sort nodes by degree	
		\State $r \gets r+1$	
		\State $minDegNode \gets$ first element from sorted list
		\State $R[minDegNode] \gets n - r$
		\State $V(G) \gets V(G)\setminus {minDegNode}$
		\State Update degrees of all \emph{$v \in N(minDegNode)$}
		\State $n \gets n-1$
		\EndWhile
		\State \Return $R$
		\EndFunction
		\State Build Graph $\vec{G}$ using $R$
	\end{algorithmic}
  \label{alg:core}
\end{algorithm}

Once the ranking of each vertex is computed (Algorithm~\ref{alg:core}), the graph is directionalized and edges \emph{u $\rightarrow$ v} are only included if and only if \emph{R(u) $<$ R(v)}. Similar to Chiba and Nishizeki's algorithm, producing the core ordering is also sequential since it requires removing a vertex from the graph in each iteration.

After the DAG is generated, kClist starts counting the cliques (Algorithm~\ref{alg:kclist}). It recursively builds subgraphs until a certain depth and then adds up the degrees of vertices in the last recursion layer. Since the DAG is not being modified any further, each vertex can be processed in parallel to reduce execution time.

% \\The pseudocode of the counting phase of kClist is as follows:

\begin{algorithm}[H]
	\caption{kClist Algorithm for Counting k-cliques}\label{countingAlg}
	\begin{algorithmic}[1]
		\State $count \gets 0$
		\State $\vec{G} \gets$ OrdCore($G$) \Comment{Directionalize input graph}
		\State Counting(\emph{k}, $\vec{G}$, \emph{n})
		\Function{Counting}{\emph{l}, $\vec{G}$, \emph{count}}
		\If{\emph{l $=$ 2}}
		\ForAll{nodes in $\vec{g}$}
		\State $count \gets count + d(node)$
		\EndFor
		\Else
		\ForAll{nodes $\in$ \emph{V($\vec{G}$)} in parallel}
		\State Build subgraph $\vec{g_{node}}$
		\State Counting(\emph{l-1}, $\vec{g_{node}}$, \emph{count})
		\EndFor
		\EndIf
		\State \Return $n$
		\EndFunction
	\end{algorithmic}
  \label{alg:kclist}
\end{algorithm}

The worst case execution time for kClist is $O(km\frac{c(G)}{2}^{k-2})$, where \emph{c(G)} is the core value, i.e. the smallest maximum out-degree generated by ordering the vertices. Since \emph{c(G) $\leq$ 2a(G)-1}, this is an improvement in the execution time over Chiba and Nishizeki for large values of \emph{k} even without considering parallelization.

Finocchi et al.~\cite{finocchi2015clique} present a scalable algorithm for counting k-cliques using the MapReduce framework. They use a degree ordering to direct the graph. More recently, Shi et al.~\cite{shi2021parallel} present Arb-Count, a new work-efficient parallel algorithm with polylogarithmic span. Aside from the core and degree ordering, they also implement the Barenboim-Elkin~\cite{barenboim2010sublogarithmic} and Goodrich-Pszona~\cite{goodrich2011external} orderings. Jain and Seshadhri present Pivoter~\cite{jain2020power}, the leading pivoting-based algorithm for k-clique counting. Their algorithm compresses the recursion tree and stores the canonical form as a Succint Clique Tree (SCT), eliminating the need for ordering the graph. Lastly, Almasri et al.~\cite{almasri2021kclique} implement ordering and pivoting-based k-clique counting algorithms on GPUs. Aside from the aforementioned dedicated solvers, various GPM frameworks~\cite{chen2020pangolin, jamshidi2020peregrine, teixeira2015arabesque, dias2019fractal} are also capable of counting cliques in large graphs. Peregrine~\cite{jamshidi2020peregrine} is a pattern-aware framework, meaning it can avoid expensive computation by only exploring relevant subgraphs. Peregrine achieves this by analyzing the input pattern (i.e., k-cliques) to develop a search plan and use it to traverse the graph. 
\section{Parallelizing the Ordering Phase}

As mentioned earlier, the counting phase requires the undirected input graph to be converted into a DAG, and kClist uses the core ordering (Algorithm~\ref{orderingAlg}) from Matula and Beck~\cite{matula1983smallest} to directionalize the graph. It ranks vertices in the graph by their effective degree. The vertices are initially sorted on the basis of their degrees and stored in a min-heap. The top element is popped and assigned the lowest rank (set to the number of vertices currently in the graph). When a vertex is removed from the graph, the degrees of its neighbors are updated and the process repeats until no vertices remain in the graph. Each edge is directed from a lower-ranked vertex to a higher-ranked vertex. This ordering is attractive because it produces the lowest maximum-degree in the graph, allowing the counting phase to theoretically perform closer to its optimal bound. However, the algorithm's vertex removal and degree update step make it inherently sequential.

% The degree-based ordering we use is similar to that described by Chen et al.\ \cite{chen2020pangolin}.
We find that using a degree-based ordering both requires less computation than a core ordering and can be parallelized. The degree ordering considers vertices' original degrees and does not require a sequential process of eliminating vertices one-by-one. To directionalize the graph based on the degree ordering (Algorithm~\ref{alg:degree}), we compare the degree of the two vertices connected by each edge. The edge is directed from the lower-degree vertex to the higher-degree vertex. If the two degrees are equal, then the edge is directed from the vertex with the lower vertex identifier to that with the higher vertex identifier. While each edge can theoretically be considered in parallel, we find vertex-parallelism is sufficient and it simplifies the code. Since our input graph is symmetrized, we only need to do a single check for the degree and vertex identifier comparison.
 % and simpler to merge new contents from \cite{beamer2015gap}

In theory, a degree-based ordering is not guaranteed to produce the lowest maximum degree, potentially increasing the work required during the counting phase. However, the time required to calculate a degree-based ordering is significantly smaller than that for computing a core ordering, especially after parallelizing. In practice, we find the degree ordering not only does not significantly slow down the counting phase, it can even accelerate it (Table~\ref{table:order}).
% The pseudocode for converting the input graph into a DAG using the degree-based ordering is as follows:
\begin{algorithm}[H]
	\caption{Building a DAG using a Degree-Based Ordering in CITRON}\label{degOrder}
	\begin{algorithmic}[1]
		\Function{GenOrdering}{\emph{G}}
		\State $degrees \gets (0, 0, ..., 0)$
		\ForAll{u $\in$ \emph{V(G)} in parallel}
		\ForAll{v $\in$ \emph{N(u)}}
		\If{$(d(u) < d(v)) \lor (d(u) = d(v) \land u < v) $}
		\State $degrees[u] \gets degrees[u] + 1$
		\EndIf
		\EndFor
		\EndFor
		\State $offsets \gets PrefixSum(degrees)$
		\ForAll{u $\in$ \emph{V(G)} in parallel}
		\ForAll{v $\in$ $\vec{N(u)}$}
		\If{$(d(u) < d(v)) \lor (d(u) = d(v) \land u < v) $}
		\State $neighbors[offsets[u]] \gets v$
		\State $offsets[u] \gets offsets[u] + 1$	
		\EndIf
		\EndFor
		\EndFor
		\State \Return $\vec{G}$
		\EndFunction
	\end{algorithmic}
  \label{alg:degree}
\end{algorithm}

\begin{table}
	\centering
	\begin{adjustbox}{width=\columnwidth}
	\begin{tabular}{lllrrrr}
		\hline
		\multicolumn{1}{c}{\textbf{Graph}} & \textbf{Abbr.} & \textbf{Description} & \textbf{$\vert$V$\vert$ (M)} & \multicolumn{1}{c}{\textbf{$\vert$E$\vert$ (M)}} & \textbf{$\vec{d}$}\\
		\hline
		As-Skitter & as & Internet topology & 1.70 & 22.19 & 13.1\\ 
		Wiki-Talk & wt & Network of Wikipedia users & 2.39 & 9.320  & 3.9\\
		Orkut & or & Social network & 3.07 & 234.37 & 76.3\\
		Cit-Patents & cp & Citations made by U.S. patents & 3.77 & 16.52 & 8.8\\
		LiveJournal & lj & Blogging community & 4.00 & 69.36 & 17.2\\
		Friendster & fr & Social network & 65.61 & 3,612.13 & 28.9\\
	\end{tabular}
	\end{adjustbox}
	\caption{Summary of the properties of input graphs used in the Evaluation. All graphs are symmetrized and taken from SNAP \cite{snapnets}.}
	\label{table:1}
\end{table} 

\begin{table*}	
	\centering
	\begin{adjustbox}{width=\textwidth}
	\begin{tabular}{lrrrr|rrrrrrr}\hline
		\multirow{3}{*}{\textbf{Graph}} 
		& \multicolumn{4}{c|}{\textbf{Core Ordering}} 
		& \multicolumn{7}{c}{\textbf{Degree Ordering}} \\             
		& \multicolumn{1}{c}{\textbf{Ordering Time}} & \textbf{Counting Time} & \textbf{Total Time} & \multicolumn{1}{c|}{\textbf{Max. Out-Degree}} & \textbf{Ordering Time} & \textbf{Speedup wrt} & \textbf{Ordering Time} & \textbf{Speedup wrt} & \textbf{Counting Time} & \multicolumn{1}{c}{\textbf{Total Time}} & \textbf{Max. Out-Degree}\\ 
		\multicolumn{1}{c}{} & & & & & \textbf{(1 thread)} & \textbf{Core Ordering} & \textbf{(96 threads)} & \textbf{Core Ordering} & & & \\\hline
		As-Skitter & 1.557 & 0.396 & 1.953 & 222 & 0.244 & 6.381 & 0.025 & 62.28 & 0.080 & \textbf{0.105} & 231\\
		Wiki-Talk & 1.409 & 0.217 & 1.626 & 260 & 0.150 & 9.393 & 0.027 & 52.185 & 0.120 & \textbf{0.147} & 340\\ 
		Orkut & 21.981 & 1.453 & 23.434 & 432 & 3.946 & 5.570 & 0.123 & 178.707 & 1.265 & \textbf{1.388} & 535\\
		Cit-Patents & 5.839 & 0.125 & 5.964 & 43 & 0.931 & 6.272 & 0.030 & 194.633 & 0.152 & \textbf{0.182} & 77\\
		LiveJournal & 6.769 & 0.264 & 7.033 & 426 & 1.006 & 6.729 & 0.038 & 178.132 & 0.244 & \textbf{0.282} & 524\\
		Friendster & 880.505 & 57.377 & 937.882 & 334 & 162.288 & 5.426 & 5.051 & 174.323 & 31.271 & \textbf{36.322} & 868\\
		\end{tabular}
	\end{adjustbox}
	\caption{Comparison between time taken to convert the input graph into a DAG using the two different orderings and the associated counting times and maximum out-degrees. The core ordering is guaranteed to produce the lowest maximum out-degree.}\label{table:order}	
\end{table*}

\section{Improving Counting Phase Scalability}
To reduce the effect of confounding factors in our experiments, we first implement the kClist algorithm within the GAP Benchmark Suite reference code (gapbs)~\cite{beamer2015gap} and we use it as a baseline. The execution time of our implementation differs only $1.27-6.40\%$ from the publicly released kClist  code (Mean: $3.72\%$, Variance: $5.98\%$). Our re-implementation effort allows us to systematically look at the impact of optimizations in isolation without also having to consider other differences in the rest of the code bases.
 
\subsection{Scaling Bottlenecks in kClist}

Our baseline implementation scales poorly for high thread counts (Figure~\ref{fig:scaling_comp}) like the released vertex-parallel implementation of kClist. Danisch et al.\ suggest load-balancing issues that arise from skewed degree distributions in sparse real-world graphs as a possible explanation for the poor scalability of their vertex-parallel algorithm. Through our analysis, we find load imbalance is at most a minor factor, and other bottlenecks hinder scalability to a much greater extent.

\begin{figure}[ht]
	\centering
	\includegraphics[width=\columnwidth]{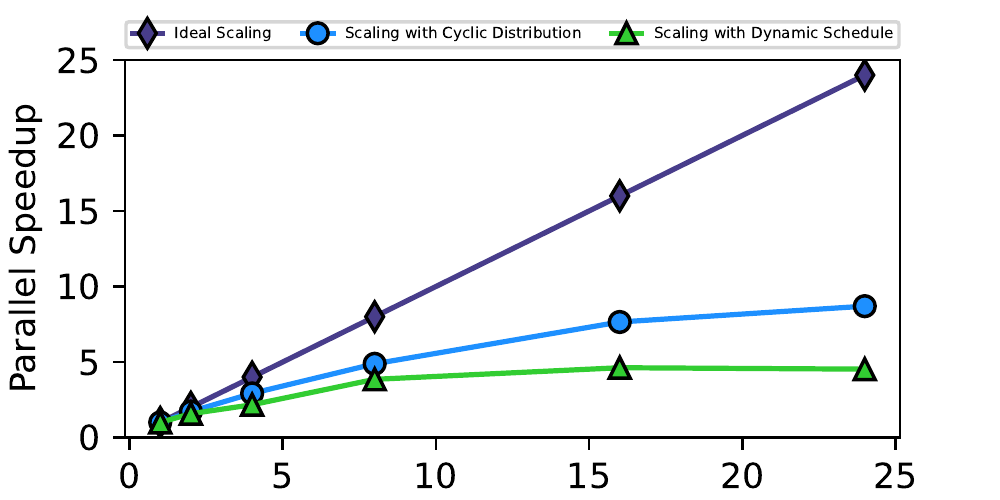}
	\caption{Parallel scaling for counting triangles in Cit-Patents using cyclic work distribution and a dynamic scheduler in the Baseline implementation.}
	\label{fig:minipage2}
\end{figure}

To analyze the impact of load imbalance, we both attempt to improve load balance as well as measure the amount of work done by each thread. We sweep various scheduling parameters such as task granularity (chunk sizes) and scheduler types (static, dynamic, cyclic), and we are not able to fully improve parallel scalability. Lumsdaine et al. show that a cyclic distribution (e.g. static schedule with chunk size of 1 in OpenMP) effectively allocates equal work among threads for triangle counting~\cite{lumsdaine2020triangle}. We test cyclic scheduling with our implementation, and we observe that parallel scaling is only moderately improved (Figure \ref{fig:minipage2}).

We measure the work performed by each thread by the number of inner loop iterations because each iteration consists of a uniform amount of work and the number of iterations while processing each vertex is dependent on the number of neighbors of that vertex. With a cyclic distribution, we are able to achieve uniform work distribution among threads. The normalized standard deviation (the ratio of the standard deviation to the mean) of work measured over 24 threads while counting triangles in Cit-Patents is 0.003.

% Lumsdaine et al.\ \cite{lumsdaine2020triangle} show that a cyclic distribution (static schedule with chunk size of 1 in OpenMP) effectively allocates equal work among the threads. We test cyclic scheduling with our implementation. We measure the work performed by each thread by the number of loop iterations because each iteration consists of a uniform amount of work and the number of iterations while processing each vertex is dependent on the number of neighbors of that vertex. With cyclic distribution, we are able to achieve uniform work distribution among threads. The normalized standard deviation (the ratio of the standard deviation to the mean) of work measured over 24 threads while counting triangles in cit-Patents is 0.003. However, we observe that parallel scaling is only moderately improved (Figure \ref{fig:minipage2}).

Shifting our analysis to bottom up, performance counter guided profiling with VTune shows memory accesses to be the bottleneck as the baseline code is 49.52\% memory bound while counting triangles on Cit-Patents and the CPU is stalled 47.6\% of the clock cycles waiting on DRAM. After our optimizations (to be detailed), according to VTune, we reduce the memory bound to 32.92\% and stalls due to DRAM to 30.6\% for the same task. Due to the success of our interventions, we can confirm that memory is a significant bottleneck of the original code.

Through our analysis, we find there are three scaling bottlenecks in our baseline implementation that closely mimics kClist. First, the DAG is built in such a way that each directed edge is included twice. This unnecessarily increases the core value, and consequently, the sizes of all the data structures associated with the graph and the subgraph. Scalability is inhibited by allocating a large subgraph structure per thread, and memory accesses quickly become a bottleneck. Second, the original implementation includes a first-level remapping of vertex identifiers changing from the interval $\left[0, \vert V(G) \vert-1\right]$ to $\left[0, c(G)-1\right]$ ($c(G)$ is the core value of the graph). We presume this remapping is done to allow the vertex identifier to function as an indexing mechanism to aid in building subgraphs, as well as improving locality. We find that the remapping step increases memory traffic and is not required to accurately count cliques. Lastly, the reference implementation updates a label vector once a neighbor is ``visited'' so that edges can be included in the subgraph for the following recursion level. This adds an extra memory access while iterating over the neighbor lists. Instead, we find that performing a set intersection of the sorted neighbor arrays is far more efficient. We find that the aforementioned implementation details place a heavy burden on memory by requiring more accesses with poor locality. By fixing these issues in the next subsection, we are able to significantly improve scalability and thus execution time.

% \subsection{Optimizations for Reducing Scaling Bottlenecks}
\subsection{Communication Reducing Optimizations}

We introduce a number of optimizations to reduce the aforementioned scaling bottlenecks. First, we build our DAG without including duplicate edges and then build compressed adjacency lists for the subgraphs in the subsequent recursion levels. We remove the relabeling step while building the first-level subgraph as well as the label-based intersection from kClist, and replace it with a set-based intersection on sorted neighbor lists. Finally, we implement certain heuristics to further reduce the amount of work done in the counting phase. 

\begin{figure}
	\centering
	\includegraphics[width=1\linewidth]{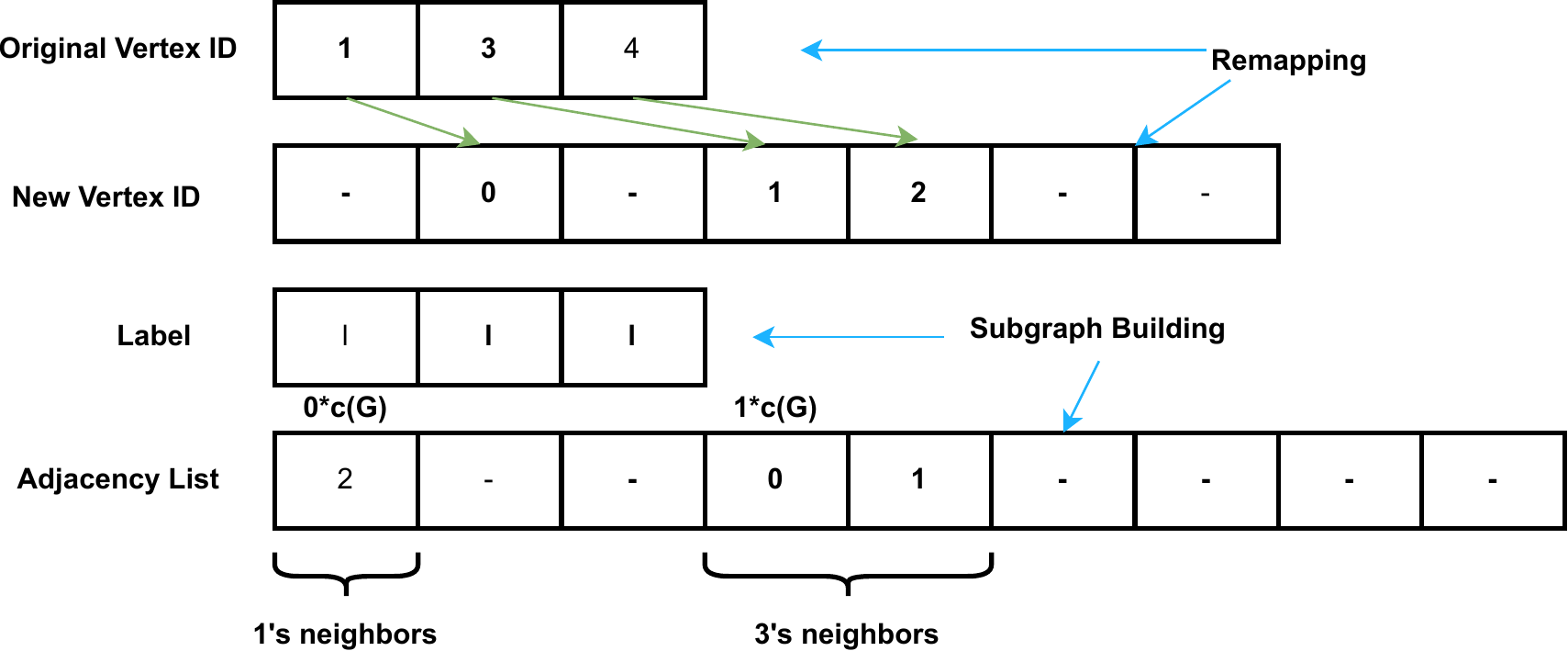}
	\caption{First level subgraph for example graph (Figure \ref{exampleGraph}) in the baseline implementation. There is also a separate array for storing the degrees of each vertex in the subgraph not pictured.}
	\label{kclist_adjlist}
\end{figure}

\paragraph{Compressing the Adjacency List}
As opposed to kClist (which contains two copies of each edge), our generated DAG contains sorted adjacency lists without duplicate neighbors. This becomes beneficial while building the subgraphs during recursion, since we can simply append each new neighbor to the end of the array. kClist pessimistically reserves $c(G)$ spaces per node in the subgraph and then starts placing the neighbors of the $i^{th}$ neighbor at index $i\times c(G)$, making the adjacency list large and sparsely filled since all entries between indices $i\times c(G)+d(\vec{i})$ and $(i+1)\times c(G)$ are unused (Figure~\ref{kclist_adjlist}). This inevitably leads to poor locality and increased cache misses. We can see the stark difference in how much memory is used for storing the first level subgraph by kClist and CITRON in Table~\ref{table:struct}. We describe both subgraph building methods in more detail in \hyperref[section:alg]{Section 5.2.2}.

\begin{figure*}
	\centering
	\includegraphics[width=1\linewidth]{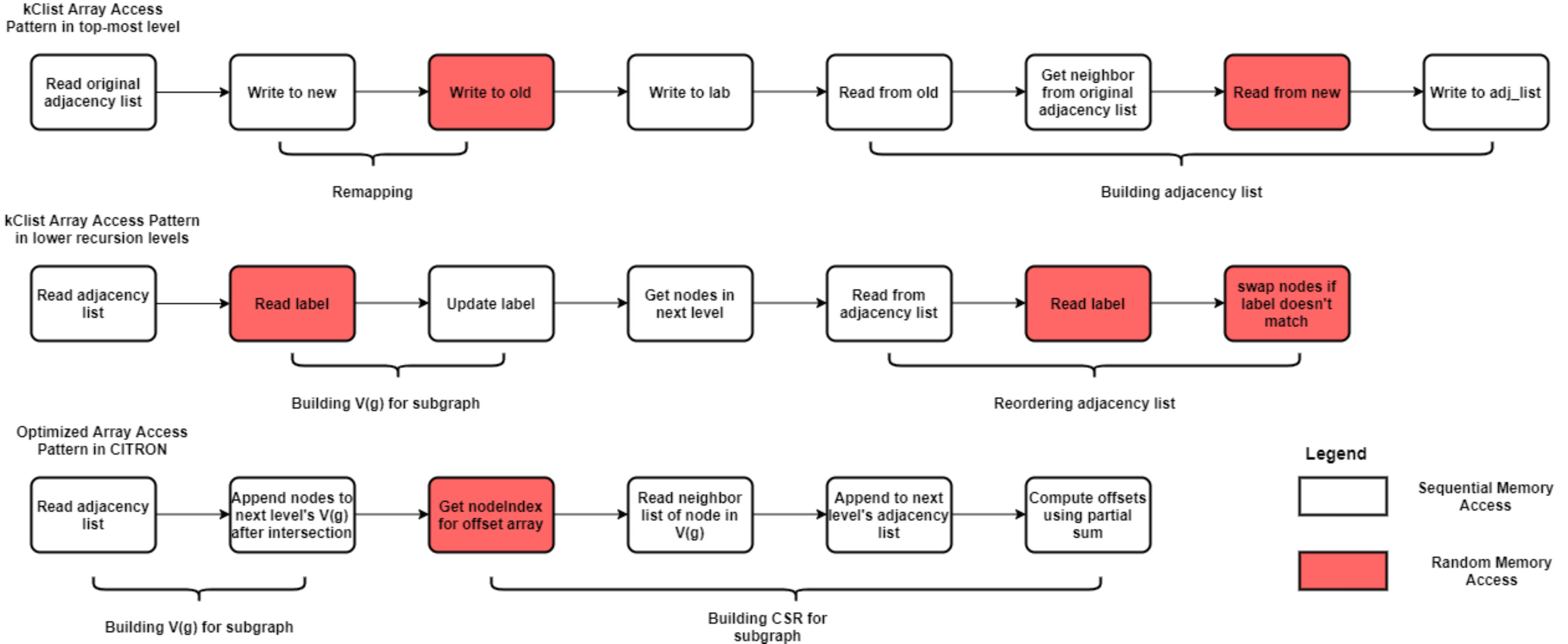}
	\caption{Memory Access Patterns for the reference kClist implementation and CITRON. The optimized version requires fewer random accesses as well as less memory accesses overall, improving performance.}
\end{figure*}

\paragraph{Reducing Memory Traffic by Removing Label Array}
\label{section:alg}
The main step of the subgraph building process is finding which edges will exist in the subgraph by intersecting each vertices' neighbor list in the original graph with $\vec{V(g)}$. KClist accomplishes this by having a single label vector with an entry per node (Algorithm~\ref{genSub}). The algorithm then loops over each node's neighbor list in the subgraph twice. Initially, there is a check to see if the label of the neighbor is equal to the current recursion level \emph{l} (Step 3). If it is equal, then the vertex has been accessed for the first time so it is added to the vertex set of the subgraph for the next level (Step 5 ) and updated to \emph{l-1} (Step 6). The second loop finds edges and updates the degrees of each node in the new vertex set. This is done by iterating over each neighbor's neighbor list and checking their label. If the label is \emph{l-1}, then an edge is present and the degree of the neighbor is incremented (Step 10).

\begin{algorithm}
	\caption{Building an Induced Subgraph in kClist}\label{genSub}
	\begin{algorithmic}[1]
		\Function{GenSubgraph}{$\vec{G}$, \emph{l}, \emph{u}}
		\State $V(\vec{g_u}) = \phi$
		\ForAll{$nodes \in$ $\vec{N(u)}$}
		\If{$lab[node] = l$}
		\State add $node$ to \emph{V($\vec{g_u}$)}
		\State $lab[node] \gets l-1$
		\EndIf
		\EndFor
		\ForAll{$node$ $\in$ \emph{V($\vec{g_u}$)}}
		\ForAll{$neighbor$ $\in$ $\vec{N(node)}$}
		\If{$lab[neighbor] = l-1$}
		\State $\vec{d(node)} \gets \vec{d(node)} + 1$
		\Else
		\State Swap nodes in adjacency list
		\EndIf
		\EndFor
		\State Reset all labels to \emph{l}
		\EndFor
		\State \Return $\vec{g_u}$
		\EndFunction
	\end{algorithmic}
\end{algorithm}

Step 11 (Algorithm~\ref{genSub}) is an optimization Danisch et al.\ \cite{danisch2018listing} mention explicitly in their paper. All the out-neighbors of a node in the next level subgraph are moved to the start of that node's neighborhood in the original adjacency list. This further increases the number of data writes required. 

One of our major contributions in CITRON is eliminating extra memory accesses for the remapping and the label-based intersection while building subgraphs (Algorithm~\ref{stdInterSub}). By having a separate adjacency list and offset array for each subgraph layer, our data is more tightly packed in contrast to the spread out, sparse structure for kClist. Instead of reading and writing from a label array (with poor locality) to check for common neighbors, we perform set intersections between \emph{V(G)} and $\vec{N(u)}$. Since our original DAG is sorted, these intersections are very efficient. Building the adjacency list only requires us to append each neighbor to the end of the existing array as we go. Since we sort the DAG in the original graph, we ensure that the individual neighbor lists in the subsequent adjacency list are also sorted without doing any extra work. Furthermore, storing each subgraph separately removes the need for swapping nodes and eliminates extra writes.  

\begin{algorithm}
	\caption{Optimized Subgraph Building in CITRON}\label{stdInterSub}
	\begin{algorithmic}[1]
		\Function{GenSubgraph}{$\vec{G}$, \emph{l}, \emph{u}}
		\State $\vec{V(g_u)} = \vec{N(u)}$
		\ForAll{node $\in$ $\vec{V(g_u)}$}
		\ForAll{neighbor $\in$ $\vec{N(node)}$}	
		\State \emph{outEdges} $\gets$ $\vec{V(g_u)}$ $\cap$ $\vec{N(neighbor)}$
		\State $adjList.append(outEdges)$
		\State $\vec{d(node)} \gets \vec{d(node)} + \vert outEdges \vert$
		\State $offsets \gets PartialSum(d)$ \Comment{for CSR}
		\EndFor
		\EndFor
		\State \Return $\vec{g_u}$
		\EndFunction
	\end{algorithmic}
\end{algorithm}

\begin{figure}
	\centering
	\includegraphics[width=1\linewidth]{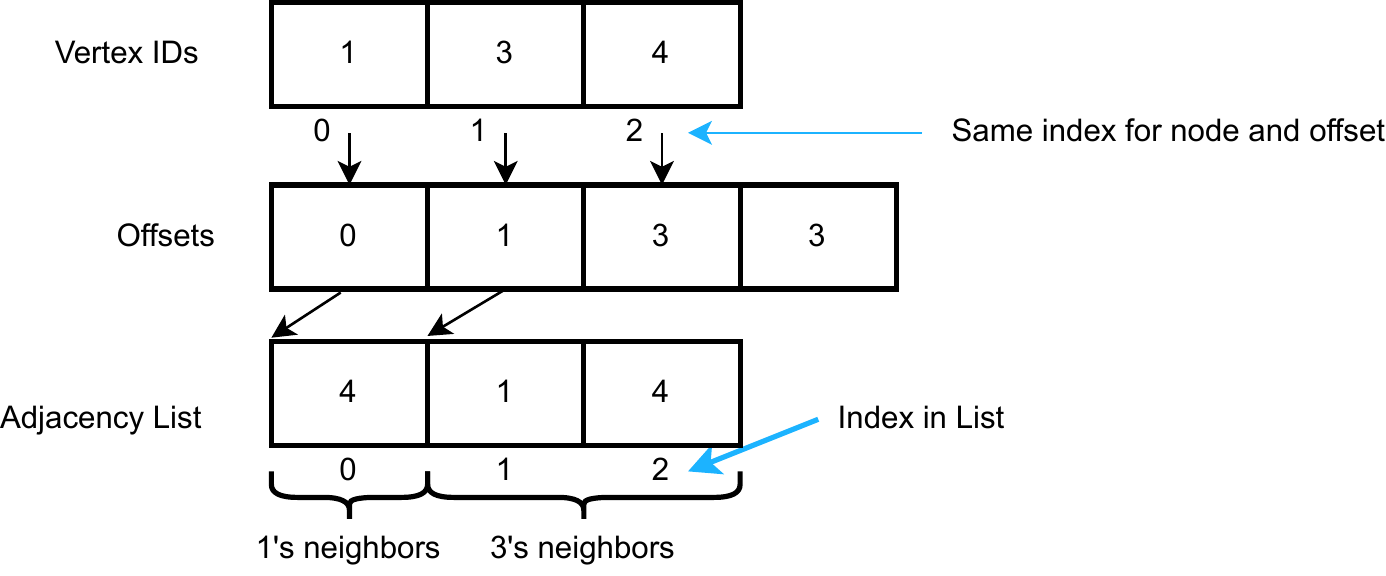}
	\caption{First level subgraph for example graph (Figure \ref{exampleGraph}) in CITRON. There is also a separate array for storing the degrees of each vertex in the subgraph not pictured.}
	\label{optimized_adjlist}
\end{figure}

To contrast the graph structures of the two approaches and how it impacts locality, we illustrate the example from Figure~\ref{exampleGraph} for both kClist (Figure~\ref{kclist_adjlist}) and CITRON (Figure~\ref{optimized_adjlist}).

% The graph structures in Figure \ref{kclist_adjlist} and Figure \ref{optimized_adjlist} illustrate the differences between how kClist and CITRON store the adjacency lists for the example graph in Figure \ref{exampleGraph} and how it impacts locality.

The adjacency list for kClist contains $c(G)$ entries reserved for each potential neighbor, i.e. $c(G)\times c(G)$ entries in total. The value of $c(G)$ for the example graph from Figure \ref{exampleGraph} is 3. For simplicity, we assume that the neighbor lists in the DAG are sorted and there are no duplicate entries. After the initial remapping phase, the new vertex identifiers are \{0, 1, 2\}.

For high-degree vertices, we can see how kClist's organization can be wasteful and require multiple accesses to fit the entire adjacency list into the cache, especially if there are more entries than a cache line can fit. KClist only does the remapping in the top-most recursion level and then swaps nodes to the front of that vertex's allocated space ($i\times c(G)$). In contrast, our approach skips the remapping and simply appends each vertex's common neighbors to the end of the array. To get the required neighbor list for the intersection, we can simply slice the adjacency list from indices \emph{Offset[nodeIndex]} to \emph{Offset[nodeIndex+1]}, where \emph{nodeIndex} is the index of the vertex in $V(G)$. Our compact adjacency list is similar to the Doubly Compressed Sparse Row (DCSR) format in CombBLAS \cite{buluc2008representation, bulucc2011combinatorial}. Additionally, a simple building process allows us to repeat it per level instead of swapping vertices and still enjoy a performance boost.

% We can get a sense of the improvements to the memory access pattern for the reference implementation and the optimization below.

Our optimizations in CITRON greatly reduces the memory access and size requirements. Table \ref{table:struct} compares the number of array accesses required to build all of the first level subgraphs and the total memory (in bytes) used by the largest subgraph in each kClist and CITRON. The kClist subgraph is of a constant size for each vertex since the array lengths are determined by the core value. In contrast, CITRON has a variable subgraph size for each vertex since the adjacency list is built by appending neighbors to the end of a list. To maintain fairness, we report the size of the largest subgraph (induced by the vertex with the highest degree).

\begin{table}	
	\centering
	\begin{adjustbox}{width=\columnwidth}
	\begin{tabular}{lrrrr}
		\hline 
		\multirow{2}{*}{\textbf{Graph}} 
		& \multicolumn{2}{c}{\textbf{Baseline}} 
		& \multicolumn{2}{c}{\textbf{CITRON}} \\             
		& \textbf{Array Accesses} & \textbf{Size of Subgraph} & \textbf{Array Accesses} & \textbf{Size of Subgraph}\\
		& \textbf{(Billions)} & \textbf{(MB)} & \textbf{(Billions)} & \textbf{(MB)}\\         
		\hline
		as & 3.179 & 13.775 & 0.565 & 0.055\\
		wt & 2.594 &  19.433 & 0.269 & 0.080\\ 
		or & 59.186 & 25.340 & 19.872 & 0.092\\
		cp & 0.562 & 30.207 & 0.305 & 0.003\\
		lj & 9.222 & 33.030 & 2.724 & 0.070 \\
		fr & 1,011.843 & 999.145 & 392.626 & 0.045\\ 
	\end{tabular}
	\end{adjustbox}
	\caption{Comparison between the number of array accesses while building all the first level subgraphs and the memory consumed for the baseline and CITRON. We measure sizes with \texttt{sizeof}}\label{table:struct}	
\end{table}

From Table \ref{table:struct}, we observe that CITRON requires fewer array accesses for all of the input graphs. We can also see that the difference in the memory sizes is quite significant, and removing the relabeling step and the label array (used for the intersections) allows us to save additional space in memory. Although the original input graph is far larger than the space reclaimed with our smaller subgraph data structures, its benefit is that it reduces the working set of memory accesses, allowing for more to fit in cache. Additionally, since each thread gets its own private copy of the subgraph struct, the memory required for storing the kClist subgraph can quickly become a bottleneck for high thread counts.
 
\paragraph{Other Optimizations}
We add a heuristic to skip processing nodes if cliques are impossible to be found. If the number of nodes in the subgraph is less than $(l-2)$ at any level in the recursion, then that subgraph will never contain a k-clique.

Finally, we further decrease the amount of work done by not building an adjacency list and computing the row offsets in the penultimate recursion layer. This is possible because we only require degrees of each node in the subgraph in the current level instead of the nodes and their neighbors in the next level. This also reduces the amount of memory required for the last level subgraph.

\section{Experimental Setup}

\subsection{Environment}
We perform our experiments on a dual-socket Intel Xeon Platinum 8260. Each socket has 24 physical cores running at 2.40GHz with two-way hyperthreading (48 threads) and 35.75MB shared L3 cache. Our system has 768 GB RAM. For the experiments, we use all thread contexts (96) unless specified otherwise. We compile with g++ (version 9.3.0) with optimization -O3 and use OpenMP. We also use Intel VTune for analyzing our optimization's improvements.   

\subsection{Graphs in this Work}
We use a variety of input graphs to evaluate the performance of our optimizations (Table~\ref{table:1}). These graphs are taken from Stanford Network Analysis Project (SNAP) \cite{snapnets}. Since clique finding is used heavily in social network analysis, we have selected commonly used graphs to make analysis more germane. All graphs are unweighted and symmetrized to initially be undirected.
% Table \ref{table:1} lists a summary of different properties of the input graphs.

\section{Evaluation}

In this section, we analyze how the two different orderings affect performance. We also compare the overall performance of our optimizations in CITRON with both vertex-parallel and edge-parallel versions of kClist, as well as a GPM framework in Peregrine~\cite{jamshidi2020peregrine}, a dedicated solver in Pivoter~\cite{jain2020power}, and a recently released code ArbCount~\cite{shi2021parallel}. All of the thread scaling experiments are performed by increasing the number of threads (and cores) from 1 to 48. All other comparisons are performed using all available threads (96). We perform at least 2 trials and record the mean for each data point. 

\subsection{Ordering Comparison}

As seen in Table \ref{table:order}, a degree-based ordering may not always produce the lowest maximum-degree value. However, its lightweight computation and parallelizability generally lead to overall lower total execution times for clique counting. In Table \ref{table:order}, we also break down the ordering and counting times for counting triangles (k=3) in each graph and see how each ordering impacts the total execution time. Both orderings are applied on our optimized code running on 96 threads and the fastest total time is denoted in bold.

The ordering time and total execution time are greatly improved with the parallel degree-based ordering. Additionally, we observe that the counting time for the degree-based ordering is also less in most cases. To understand this, we make a simple model for algorithmic work and animate it with data from our executions. For triangles, the total work is given by the sum  of the product of each vertex's degree with the sum of each of it's neighbor's degrees as the algorithm traverses those edges ($\sum_{u \in V(G)}^{} (d_u.\sum_{v \in N(\vec{u})}^{} d_v)$). The degree-based ordering produces a smaller value for work than the core value in most cases. As k gets larger, core ordering becomes more beneficial. Increasing k increases the max recursive depth, and the overall execution becomes more dominated by the highest degree nodes.
% As k gets larger, the fan-out of computing subgraphs increases with the degree. Thus, the core ordering leads to better performance in the counting phase for higher k.
 
\subsection{Parallel Scaling of the Counting Phase}

\begin{figure*}
	\centering
	\includegraphics[width=1\textwidth]{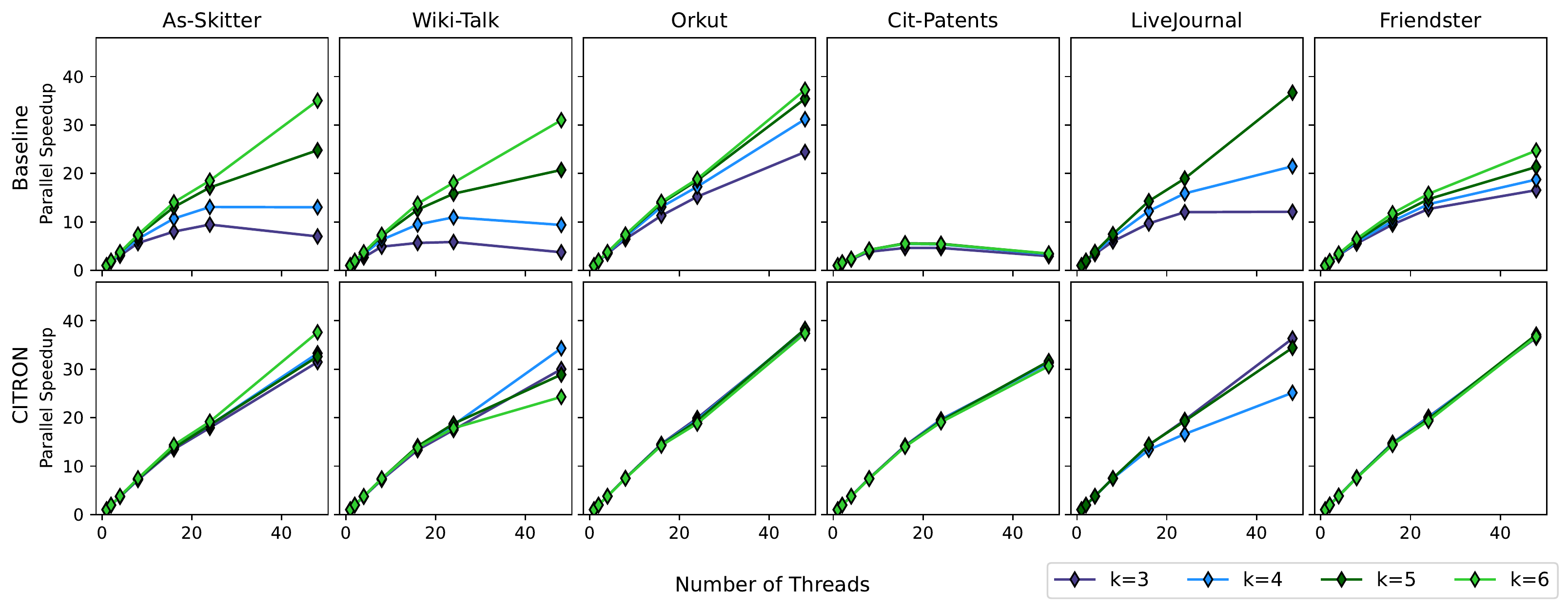}

	\caption{Comparing parallel scaling between the baseline and CITRON for counting cliques of various sizes in all input graphs. All the subplots on the top row are for the baseline implementation and the all the subplots on the bottom row are for CITRON. The graphs are: (left to right) As-Skitter, Wiki-Talk, Orkut, Cit-Patents, LiveJournal, Friendster. At least two trials were completed and the mean was taken to account for variance. Scaling for 6-cliques on LiveJournal is not included as the single-thread runtime is very high. }\label{fig:scaling_comp}
\end{figure*}

To compare the improvement in parallel scaling of our optimizations with our baseline implementation, we count 3,4,5 and 6-cliques on the input graphs on increasing thread counts up to 48 threads (Figure~\ref{fig:scaling_comp}). We do not see major improvement using hyperthreading, so we report data using 48 threads on 2 sockets and 1-24 threads on one socket. Each plot shows the parallel speedup of the counting phase relative to single thread of that implementation.
% All the plots on the top in Figure \ref{fig:scaling_comp} are for the baseline implementation and those on the bottom are for CITRON.

In the baseline scaling plots (Figure~\ref{fig:scaling_comp}), we observe that scaling improves as k increases. We believe the amortization of scheduling overhead causes this phenomenon since more work is required to count larger cliques than smaller cliques. We can see that our optimizations in CITRON reduce this effect. Especially in Orkut and cit-Patents, all k values scale equally well. Most notably, our optimizations perform extremely well compared to kClist on cit-Patents, where we achieve near-linear scaling.
% We also observe that both kClist and CITRON achieve poor scaling on Road. Road only has very few triangles (k = 3) and 4-cliques. Thus, the runtime for the counting phase is very small and we suspect there isn't enough work to allocate to multiple threads.

\subsection{Total execution time Comparison}

Table \ref{table:3} shows the overall execution times for counting cliques of various sizes on all the input graphs for each implementation. We use the released implementation of kClist rather than our baseline for fairness. We were unable to convert As-Skitter into the required graph format for Peregrine. After repeated effort, we are unable to use KClist's Edge-Parallel implementation to count triangles on any graph. Peregrine is unable to count cliques of any size on Friendster as it times out. 

\begin{table*}
	\centering
	\begin{tabular}{llrrrrrr}
		\hline
		\multicolumn{1}{c}{\textbf{Graph}} & \textbf{Algorithm} & \textbf{k=3} & \multicolumn{1}{c}{\textbf{k=4}} & \textbf{k=5} & \textbf{k=6} & \textbf{k=7} & 
		\multicolumn{1}{c}{\textbf{k=8}}\\
		\hline
		
		\textbf{As-Skitter} & CITRON & \emph{0.109} & \emph{0.159} & \emph{0.410} & \emph{2.383} & 20.754 & 187.908\\
		& kClist Vertex-Parallel & 2.358 & 2.380 & 2.636 & 4.508 & \emph{18.438} & 115.348\\ 
		& kClist Edge-Parallel & - & 4.273 & 4.502 & 6.368 & 20.240 & 111.984\\ 
		& ArbCount & \textbf{0.0470} & \textbf{0.073} & \textbf{0.127} & \textbf{0.635} & \textbf{5.280} & 39.699\\
		& Pivoter & 30.000 & 32.667 & 35.333 & 37.000 & 39.000 & \textbf{39.667}\\
		\hline
		
		\textbf{Wiki-Talk} & CITRON & \emph{0.217} & \emph{0.240} & \emph{0.399} & \emph{1.442} & 6.152 & 22.925\\
		& kClist Vertex-Parallel & 1.638 & 1.653 & 1.788 & 2.383 & \emph{4.430} & \emph{10.131}\\ 
		& kClist Edge-Parallel & - & 2.792 & 2.935 & 3.494 & 5.534 & 11.048\\ 
		& ArbCount & \textbf{0.042} & \textbf{0.045} & \textbf{0.092} & \textbf{0.468} & \textbf{2.288} & \textbf{8.747}\\
		& Peregrine & 0.928 & 10.405 & 68.781 & 719.048 & $>$1h & $>$1h\\
		& Pivoter & 32.667 & 36.000 & 40.667 & 43.333 & 44.000 & 46.000\\
		\hline
		
		\textbf{Orkut}	& CITRON & \emph{1.437} & \emph{3.163} & \emph{10.043} & \emph{39.441} & \emph{170.447} & 763.592\\
		& kClist Vertex-Parallel & 44.442 & 46.318 & 52.980 & 78.320 & 184.128 & 624.370\\ 
		& kClist Edge-Parallel & - & 77.152 & 83.748 & 110.199 & 218.507 & 672.369\\ 
		& ArbCount & \textbf{1.201} & \textbf{1.614} & \textbf{2.863} & \textbf{8.694} & \textbf{33.278} & \textbf{133.695}\\ 
		& Peregrine & 7.362 & 144.532 & 1587.830 & $>$1h & $>$1h & $>$1h\\
		& Pivoter & 227.333 & 308.667 & 400.667 & 481.333 & 525.333 & \emph{583.333}\\
		\hline
		
		\textbf{Cit-Patents} & CITRON & \emph{0.187} & \textbf{0.197} & \emph{0.206} & \emph{0.216} & \emph{0.231} & \emph{0.243}\\
		& kClist Vertex-Parallel & 9.918 & 9.867 & 9.876 & 9.894 & 9.904 & 9.868\\ 
		& kClist Edge-Parallel & - & 14.128 & 14.108 & 14.136 & 14.098 & 14.114\\ 
		& ArbCount & \textbf{0.122} & 0.204 & \textbf{0.190} & \textbf{0.181} & \textbf{0.174} & \textbf{0.164}\\
		& Peregrine & 0.349 & 0.419 & 0.415 & 0.512 & 1.070 & 1.599\\
		& Pivoter & 62.667 & 63.000 & 63.000 & 63.333 & 63.667 & 63.000 \\
		\hline

		\textbf{LiveJournal} & CITRON & \emph{0.379} & \emph{0.896} & 21.475 & 980.062 & $>$1h & $>$1h\\
		& kClist Vertex-Parallel & 9.400 & 9.666 & \emph{17.107} & 378.124 & $>$1h & $>$1h\\ 
		& kClist Edge-Parallel & - & 15.212 & 23.338 & \emph{371.418} & $>$1h & $>$1h\\ 
		& ArbCount & \textbf{0.204} & \textbf{0.416} & \textbf{5.587} & \textbf{256.241} & $>$1h & $>$1h\\ 
		& Peregrine & 0.537 & 5.474 & 218.286 & $>$1h & $>$1h & $>$1h\\
		& Pivoter & 118.500 & 299.500 & 1500.000 & $>$1h & $>$1h & $>$1h\\
		\hline

		\textbf{Friendster} & CITRON & \emph{37.790} & \textbf{46.825} & \textbf{65.201} & \emph{106.352} & \emph{258.574} & 1765.970\\
		& Baseline$^*$ & 1275.195 & 1278.400 & 1281.445 & 1292.870 & 1311.890 & \emph{1354.710}\\
		& ArbCount & \textbf{31.288} & 70.010 & 70.817 & \textbf{74.418} & 93.549 & \textbf{385.024}\\
		& Pivoter & 4314.500 & 4433.500 & 4489.500 & 4554.500 & 4537.500 & 4556.500\\
		
	\end{tabular}
	\caption{Summary of total execution time for counting cliques using kClist \cite{danisch2018listing}, CITRON, Peregrine \cite{jamshidi2020peregrine} and Pivoter \cite{jain2020power}. Each algorithm was executed using 96 threads on the same machine under the same conditions. The best execution time is denoted in bold. The best execution time other than ArbCount is italicized. $^*$Both Node Parallel and Edge Parallel version of kClist segfaulted for Friendster so we report the values for our baseline implementation which has nearly identical performance to the kClist Node Parallel implementation.}
	\label{table:3}
\end{table*}

\begin{figure*}
	\centering
	\includegraphics[width=1\linewidth]{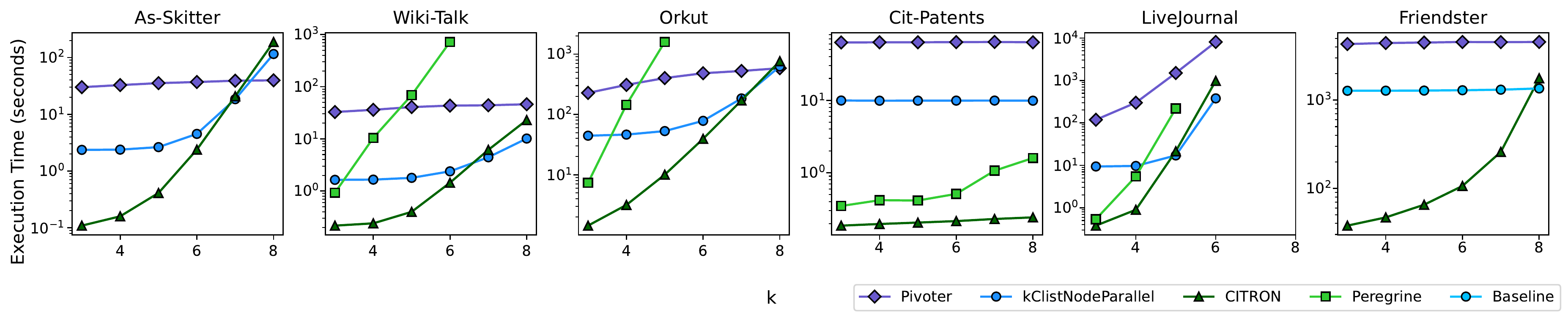}
	\caption{Plots in linlog scale comparing the total execution time required for counting cliques of different sizes on the input graphs using kClist \cite{danisch2018listing}, CITRON, Peregrine \cite{jamshidi2020peregrine} and Pivoter \cite{jain2020power}. kClistEdgeParallel is not included as kClistNodeParallel outperforms it for every graph. As kClist segfaults on Friendster, we include the times of our baseline implementation. On LiveJournal, none of the frameworks are unable to count 7, 8-cliques within an hour.}
	\label{ord_scaling}
\end{figure*}

Our speedups range from $0.39-53.04\times$ (Geometric Mean: $6.19\times$) over kClist, $1.42-498.65\times$ (Geometric Mean: $10.71\times$) over Peregrine and $0.21-335.12\times$ (Geometric Mean: $45.08\times$) over Pivoter. CITRON performs best for modest values of k (Figure~\ref{ord_scaling}). The newly released ArbCount outperforms all frameworks most of the time.

We observe that a degree ordering greatly improves the execution time for smaller k values as the overhead of computing the core ordering is very large compared to the counting time. However, as counting time increases with the clique size, this overhead becomes amortized and the core ordering leads to a smaller execution time. This is consistent with Shi et al.'s findings \cite{shi2021parallel}. 

We suspect that load imbalance likely becomes an issue with our optimizations for higher k values. With a degree-based ordering, the larger maximum degree increases the computational fan-out for building subgraphs in deeper recursions. Our asymptotic analysis also shows that the execution time of the counting phase also grows exponentially for increasing k. We parallelize a task size of 1 (i.e.\ each vertex). Initial analysis shows counting cliques in the induced subgraphs of large degree vertices require a significant fraction of the total execution time. Thus, it may be beneficial to parallelize subgraph building or further recursion levels for large degree vertices.

\section{Conclusion}
CITRON optimizes kClist's vertex-parallel algorithm by optimizing memory communication and increasing parallelization, both proven techniques to improve graph algorithm performance~\cite{Beamer-thesis}. CITRON reduces communication between the CPU and memory by more efficiently storing and accessing the subgraph structures in each recursion level of the counting phase. We are able to significantly improve parallel scaling and the time required to count cliques using this optimization. We also use a parallel, lightweight degree-based ordering to produce a DAG in the ordering phase. Even though we lose some efficiency in the counting phase by using the degree-based ordering instead of a core ordering, we observe improvements in total execution time because the degree ordering is faster to compute. CITRON achieves speedups of $18.16-208.27\times$ (Geometric Mean: $102.436\times$) in the ordering phase and speedups of $1.1-2.6\times$ (Geometric Mean: 1.798$\times$) in the counting phase, resulting in total speedups of $14.37-39.37\times$ (Geometric Mean: $18.517\times$) over kClist while counting triangles.

While we focus on counting in this work, we are easily able to enumerate each clique with a few simple changes to our code. Since more research is in the field of Graph Pattern Mining, we hope that our optimization can be used as a performance benchmark to compare the performance of new clique counting algorithms.

\FloatBarrier
\section*{Acknowledgment}

\bibliographystyle{IEEEtran}
\bibliography{references}

% Generated by IEEEtran.bst, version: 1.12 (2007/01/11)
\begin{thebibliography}{10}
\providecommand{\url}[1]{#1}
\csname url@samestyle\endcsname
\providecommand{\newblock}{\relax}
\providecommand{\bibinfo}[2]{#2}
\providecommand{\BIBentrySTDinterwordspacing}{\spaceskip=0pt\relax}
\providecommand{\BIBentryALTinterwordstretchfactor}{4}
\providecommand{\BIBentryALTinterwordspacing}{\spaceskip=\fontdimen2\font plus
\BIBentryALTinterwordstretchfactor\fontdimen3\font minus
  \fontdimen4\font\relax}
\providecommand{\BIBforeignlanguage}[2]{{%
\expandafter\ifx\csname l@#1\endcsname\relax
\typeout{** WARNING: IEEEtran.bst: No hyphenation pattern has been}%
\typeout{** loaded for the language `#1'. Using the pattern for}%
\typeout{** the default language instead.}%
\else
\language=\csname l@#1\endcsname
\fi
#2}}
\providecommand{\BIBdecl}{\relax}
\BIBdecl

\bibitem{shi2021parallel}
J.~Shi, L.~Dhulipala, and J.~Shun, ``Parallel clique counting and peeling
  algorithms,'' in \emph{SIAM Conference on Applied and Computational Discrete
  Algorithms (ACDA21)}.\hskip 1em plus 0.5em minus 0.4em\relax SIAM, 2021, pp.
  135--146.

\bibitem{6249683}
E.~Gregori, L.~Lenzini, and S.~Mainardi, ``Parallel k-clique community
  detection on large-scale networks,'' \emph{IEEE Transactions on Parallel and
  Distributed Systems}, vol.~24, no.~8, pp. 1651--1660, 2013.

\bibitem{palla2005uncovering}
G.~Palla, I.~Der{\'e}nyi, I.~Farkas, and T.~Vicsek, ``Uncovering the
  overlapping community structure of complex networks in nature and society,''
  \emph{nature}, vol. 435, no. 7043, pp. 814--818, 2005.

\bibitem{7117352}
F.~Hao, G.~Min, Z.~Pei, D.-S. Park, and L.~T. Yang, ``$ k $-clique community
  detection in social networks based on formal concept analysis,'' \emph{IEEE
  Systems Journal}, vol.~11, no.~1, pp. 250--259, 2017.

\bibitem{fang2019efficient}
Y.~Fang, K.~Yu, R.~Cheng, L.~V. Lakshmanan, and X.~Lin, ``Efficient algorithms
  for densest subgraph discovery,'' \emph{arXiv preprint arXiv:1906.00341},
  2019.

\bibitem{4811845}
L.~Pan and E.~E. .~Santos, ``An anytime-anywhere approach for maximal clique
  enumeration in social network analysis,'' in \emph{2008 IEEE International
  Conference on Systems, Man and Cybernetics}, 2008, pp. 3529--3535.

\bibitem{rossi2015parallel}
R.~A. Rossi, D.~F. Gleich, and A.~H. Gebremedhin, ``Parallel maximum clique
  algorithms with applications to network analysis,'' \emph{SIAM Journal on
  Scientific Computing}, vol.~37, no.~5, pp. C589--C616, 2015.

\bibitem{dietrec}
S.~Manoharan and Sathish, ``Patient diet recommendation system using k clique
  and deep learning classifiers,'' \emph{Journal of Artificial Intelligence and
  Capsule Networks}, vol.~2, no.~2, pp. 121--130, 2020.

\bibitem{movierec}
K.~X. P.~Vilakone and D.~Park, ``Personalized movie recommendation system
  combining data mining with the k-clique method,'' \emph{Journal of
  Information Processing Systems}, vol.~15, no.~5, pp. 1141--1155, Oct. 2019.

\bibitem{variant}
\BIBentryALTinterwordspacing
T.~Marschall, I.~G. Costa, S.~Canzar, M.~Bauer, G.~W. Klau, A.~Schliep, and
  A.~Schönhuth, ``{CLEVER: clique-enumerating variant finder},''
  \emph{Bioinformatics}, vol.~28, no.~22, pp. 2875--2882, 10 2012. [Online].
  Available: \url{https://doi.org/10.1093/bioinformatics/bts566}
\BIBentrySTDinterwordspacing

\bibitem{database}
E.~T. T.~Matsunaga, C.~Yonemori and M.~Muramatsu, ``Clique-based data mining
  for related genes in a biomedical database,'' \emph{BMC Bioinformatics},
  vol.~10, no. 205, 2009.

\bibitem{ppi}
K.~C.~D. Bahadur, T.~Akutsu, E.~Tomita, T.~Seki, and A.~Fujiyama, ``Point
  matching under non-uniform distortions and protein side chain packing based
  on an efficient maximum clique algorithm,'' \emph{Genome Informatics},
  vol.~13, pp. 143--152, 2002.

\bibitem{chen2020pangolin}
X.~Chen, R.~Dathathri, G.~Gill, and K.~Pingali, ``Pangolin: An efficient and
  flexible graph mining system on cpu and gpu,'' \emph{Proceedings of the VLDB
  Endowment}, vol.~13, no.~8, pp. 1190--1205, 2020.

\bibitem{jamshidi2020peregrine}
K.~Jamshidi, R.~Mahadasa, and K.~Vora, ``Peregrine: a pattern-aware graph
  mining system,'' in \emph{Proceedings of the Fifteenth European Conference on
  Computer Systems}, 2020, pp. 1--16.

\bibitem{teixeira2015arabesque}
C.~H. Teixeira, A.~J. Fonseca, M.~Serafini, G.~Siganos, M.~J. Zaki, and
  A.~Aboulnaga, ``Arabesque: a system for distributed graph mining,'' in
  \emph{Proceedings of the 25th Symposium on Operating Systems Principles},
  2015, pp. 425--440.

\bibitem{dias2019fractal}
V.~Dias, C.~H. Teixeira, D.~Guedes, W.~Meira, and S.~Parthasarathy, ``Fractal:
  A general-purpose graph pattern mining system,'' in \emph{Proceedings of the
  2019 International Conference on Management of Data}, 2019, pp. 1357--1374.

\bibitem{danisch2018listing}
M.~Danisch, O.~Balalau, and M.~Sozio, ``Listing k-cliques in sparse real-world
  graphs,'' in \emph{Proceedings of the 2018 World Wide Web Conference}, 2018,
  pp. 589--598.

\bibitem{jain2020power}
S.~Jain and C.~Seshadhri, ``The power of pivoting for exact clique counting,''
  in \emph{Proceedings of the 13th International Conference on Web Search and
  Data Mining}, 2020, pp. 268--276.

\bibitem{matula1983smallest}
D.~W. Matula and L.~L. Beck, ``Smallest-last ordering and clustering and graph
  coloring algorithms,'' \emph{Journal of the ACM (JACM)}, vol.~30, no.~3, pp.
  417--427, 1983.

\bibitem{chiba1985arboricity}
N.~Chiba and T.~Nishizeki, ``Arboricity and subgraph listing algorithms,''
  \emph{SIAM Journal on computing}, vol.~14, no.~1, pp. 210--223, 1985.

\bibitem{finocchi2015clique}
I.~Finocchi, M.~Finocchi, and E.~G. Fusco, ``Clique counting in mapreduce:
  Algorithms and experiments,'' \emph{Journal of Experimental Algorithmics
  (JEA)}, vol.~20, pp. 1--20, 2015.

\bibitem{barenboim2010sublogarithmic}
L.~Barenboim and M.~Elkin, ``Sublogarithmic distributed mis algorithm for
  sparse graphs using nash-williams decomposition,'' \emph{Distributed
  Computing}, vol.~22, no. 5-6, pp. 363--379, 2010.

\bibitem{goodrich2011external}
M.~T. Goodrich and P.~Pszona, ``External-memory network analysis algorithms for
  naturally sparse graphs,'' in \emph{European Symposium on Algorithms}.\hskip
  1em plus 0.5em minus 0.4em\relax Springer, 2011, pp. 664--676.

\bibitem{almasri2021kclique}
M.~Almasri, I.~E. Hajj, R.~Nagi, J.~Xiong, and W.~mei Hwu, ``K-clique counting
  on gpus,'' 2021.

\bibitem{snapnets}
J.~Leskovec and A.~Krevl, ``{SNAP Datasets}: {Stanford} large network dataset
  collection,'' \url{http://snap.stanford.edu/data}, Jun. 2014.

\bibitem{beamer2015gap}
S.~Beamer, K.~Asanovi{\'c}, and D.~Patterson, ``The gap benchmark suite,''
  \emph{arXiv preprint arXiv:1508.03619}, 2015.

\bibitem{lumsdaine2020triangle}
A.~Lumsdaine, L.~Dalessandro, K.~Deweese, J.~Firoz, and S.~McMillan, ``Triangle
  counting with cyclic distributions,'' in \emph{2020 IEEE High Performance
  Extreme Computing Conference (HPEC)}.\hskip 1em plus 0.5em minus 0.4em\relax
  IEEE, 2020, pp. 1--8.

\bibitem{buluc2008representation}
A.~Buluc and J.~R. Gilbert, ``On the representation and multiplication of
  hypersparse matrices,'' in \emph{2008 IEEE International Symposium on
  Parallel and Distributed Processing}.\hskip 1em plus 0.5em minus 0.4em\relax
  IEEE, 2008, pp. 1--11.

\bibitem{bulucc2011combinatorial}
A.~Bulu{\c{c}} and J.~R. Gilbert, ``The combinatorial blas: Design,
  implementation, and applications,'' \emph{The International Journal of High
  Performance Computing Applications}, vol.~25, no.~4, pp. 496--509, 2011.

\bibitem{Beamer-thesis}
S.~Beamer, ``Understanding and improving graph algorithm performance,'' Ph.D.
  dissertation, University of California, Berkeley, 2016.

\end{thebibliography}

\end{document}